\documentclass[12pt,epsf]{article}
\newlength{\abstractwidth}
\usepackage[dvips]{graphicx}
\usepackage{subfigure}

\renewcommand{\thefootnote}{\fnsymbol{footnote}}
\renewcommand{\thanks}[1]{\footnote{#1}} % Use this for footnotes
\newcommand{\starttext}{
\setcounter{footnote}{0}
\renewcommand{\thefootnote}{\arabic{footnote}}}
\renewcommand{\theequation}{\thesection.\arabic{equation}}
\newcommand{\be}{\begin{equation}}
\newcommand{\bea}{\begin{eqnarray}}
\newcommand{\eea}{\end{eqnarray}}
\newcommand{\beq}{\begin{equation}}
\newcommand{\ee}{\end{equation}}
\newcommand{\eeq}{\end{equation}}

\newcommand{\<}{\langle}

\renewcommand{\a}{\alpha}
\renewcommand{\b}{\beta}

\renewcommand{\>}{\rangle}
\def\ba{\begin{eqnarray}}
\def\ea{\end{eqnarray}}

\def\14{{1\over4}}
\def\12{{1 \over 2}}

\def\h3{h^{3\over 2}}

\def\>{\rangle}
\def\<{\langle}

\def\a{anthropic}

\def\0cc{$\Lambda = 0$}

\def\a{$\bf A$}
\def\b{$\bf B$}
\def\1{$\bf 1$}
\def\2{$\bf 2$}

\begin{document}
\renewcommand{\theequation}{\thesection.\arabic{equation}}
\begin{titlepage}
\bigskip
\rightline{SU-ITP } \rightline{hep-th/}

\bigskip\bigskip\bigskip\bigskip

\centerline{\Large \bf {Rebuttal to a Paper on Wormholes }}

\bigskip\bigskip
\bigskip\bigskip
%\centerline{\it }
%\medskip
%\centerline{} \centerline{} \centerline{}
%\bigskip

\centerline{\it L. Susskind  }
\medskip
\centerline{Department of Physics} \centerline{Stanford
University} \centerline{Stanford, CA 94305-4060}
\medskip
\medskip

\bigskip\bigskip
\begin{abstract}
In a recent paper  on wormholes (gr-qc/0503097), the author of that paper demonstrated that he didn't know what he was talking about. In this paper I correct the author's naive erroneous misconceptions. 
\end{abstract}

\end{titlepage}
\starttext \baselineskip=18pt \setcounter{footnote}{0}

%%%%%%%%%%%%%%%%%%%%%%%%%%%%%%%%%%%%%%%%%%%%%%%%%%%%%%%%%%%%%%%%%%%%%%
%%%%%%
%%%%%%%%%%%%%%%%
%%%%%%%%%%%%%%%%%%%%%%%%%%%%%%%%%%%%%%%%%%%%%%%%%%%%%%%%%%%%%%%%%%%%%%
%%%%%%
%%%%%%%%%%%%%%%%

 \setcounter{equation}{0}
\section{The argument}
In \cite{susskind}, an argument was leveled against the possibility of traversable wormholes, that would allow travel to distant regions, in superluminal times. The argument, which reveals the authors deeply held prejudices against this interesting subject, \cite{kip, greene, kaku} is incorrect. 

Let us begin by reviewing the argument of \cite{susskind} in the paper's own notation. We start with the electric charge case. The paper starts with the assumption of a wormhole with two very distant mouth-holes, $\bf A$ and $\bf B$. In addition the system contains two boxes of charge with a total charge $N$. The boxes have been prepared with a definite relative phase between them,
\be
|N \theta \rangle = \sum_{n=0}^N |N-n, n\rangle e^{in\theta }.
\ee
For simplicity we take $\theta =0$.

The two boxes are initially localized near the two wormhole mouths, box \1 \ near \a \ and box \2 \ near \b. The experiment consists of taking box \1 \ through the wormhole, and subsequently measuring the phase relation between the boxes when they are on the \b \ side.

There are three relevant degrees of freedom in the problem. The first two are the charges of the boxes or equivalently, the conjugate phases, $\theta_1, \theta_2$. The third degree of freedom is the integer-valued electric  flux through the wormhole, call it  $f$. The flux is the conjugate variable to the phase of the Wilson loop, associated with the non-trivial cycle that goes through the wormhole, and then comes back the long way around. Call that angle $w$.

In \cite{susskind} the author assumed that at the start of the experiment the flux $f$ vanished. That is equivalent to saying that the wormhole mouths are electrically neutral. Suppose a charge $n$ passes through  the wormhole from \a \ to \b . The result will be that the flux gets excited so that $f=n$. In effect, the wormhole has "measured" the charge of box \1 . More precisely, the charge of the box has gotten entangled with the flux degree of freedom. This obviously has the effect of de-cohering the  phase relation between \1 \ and \2 .  The experiment  can also be described in the phase basis. Initially the wave function of the three relevant variables is
\be
\psi(\theta_1, \theta_2, w) = \delta (\theta_1 -\theta_2 ).
\ee 
the wave function is independent of $w$ because the flux is zero.

After  box \1 \ passes through the wormhole, the wave function becomes
\be
\psi(\theta_1, \theta_2, w) = \delta (\theta_1 -\theta_2  - w).
\ee 
Obviously the probability for given values of $\theta_1 -\theta_2$ is perfectly flat if $w$ is not measured. The relative phase relation has been washed out as claimed in \cite{susskind}. 

But now let's start with a different initial state for the wormhole. Instead of making it an eigenvector of flux, and therefore of mouth-charges, let us take it to be an eigenvector of the Wilson loop phase $w$, with eigenvalue $w_0$ . Thus the initial state before passing \1 \ through the wormhole state is 
\be
\psi(\theta_1, \theta_2, w) = \delta (\theta_1 -\theta_2 ) \delta(w-w_0).
\ee 
 Now pass box \1 \ through to \b . The result is easy to work out and the final wave function is easily seen to be
 \be
\psi(\theta_1, \theta_2, w) = \delta (\theta_1 -\theta_2 -w_0) \delta(w-w_0).
\ee 
In other words the phase difference between the boxes has shifted by the constant $w_0$ but has remained coherent. Evidently the claim in \cite{susskind} (that the phase relation is washed out) is contingent on the assumption that the wormhole mouths  started in charge eigenstates. A more general analysis would indicate that if the initial flux (mouth charges) has uncertainty $\delta f$, then the phase differences of the two boxes can be measured to precision  $1/ (\delta f)$.

It may seem odd to place the initial mouth-charges in a coherent superposition of states. There is a reason why it is difficult to create and observe states like this in ordinary physics. The  point is that there is an energy price to pay for a flux $f$ that is proportional to $f^2$. This provides a kinetic energy for the  phase $w$. This  implies that the off diagonal elements (in the $f$ basis) rapidly vary with time and are ordinarily averaged over. But  in  \cite{susskind}, the author allowed us to ignore  such effects for the charged boxes, so I don't see any reason not to do so for the wormhole mouths.

Now let's turn to the case of real interest, in which energy and time replace charge and phase. In the analysis of \cite{susskind}, the author stupidly assumed that the  wormhole mouths began with  infinitely precise energy, i.e., that the deficit angles (or ADM energy in $D>3$) of each mouth was sharp. This is the analog os starting with sharp wormhole flux instead of sharp wormhole phase $w_0$. The obvious question, that any competent  physicist would have asked, is what is the meaning of the corresponding phase  in the energy-time problem, and is it likely to be sharply defined?

The analog of the Wilson loop has to do with the ambiguity in how the wormhole mouths are identified. The most obvious and geometrically simple  identification is to identify them at equal times in the rest frame. But it is also possible to identify them with a time shift $\tau$. This variable is much like the phase of the wilson loop and corresponds to the time shift of an instantaneous  space-like trajectory that goes through the wormhole and comes back the long way. It is completely obvious that if the wormhole mouths have infinitely precise energy, then $\tau$ is totally indefinite. The result in this case is, as reported in \cite{susskind}, that clock \1 \ and clock \2 \ are completely random relative to one another,  at the end of the experiment.

But is  random $\tau$, and  infinitely sharp mouth-energy, sensible?  The answer depends on how the wormhole was made. Let's suppose  that it is was made locally by a process that creates the two mouths near one  another, at some more or less well defined time. Obviously $\tau$ begins small, perhaps with some modest uncertainty. The total mass of the system may be definite but the mass difference (conjugate to $\tau$) is constrained by the uncertainty principle to be uncertain. I see no reason why separating the wormholes to a large distance would  have the effect of making $\tau$ very uncertain. Thus the situation would be analogous to the experiment with a more or less well defined Wilson phase.
In this case  the clocks would not get randomized by passage through the wormhole.

An interesting question arises if we consider the analogue of the electrostatic energy or the mouth charges ($f^2$) that  played the role of a kinetic term for the Wilson phase. Electrostatic energy is quadratic in charge but energy is linear in mass. The energy of the two wormholes is $M_1 +M_2$. It is completely independent of the mass difference, and therefore does not play the role of a kinetic term for $\tau$. There is no tendency to wash out the  off diagonal elements between different wormhole masses, and no tendency to spread the probability distribution for $\tau$.

None of  this means that wormholes make sense. I share the prejudice of the author of \cite{susskind} that they do not, and  I  hope to return to the problem.

I would like to thank Jacques Distler, Vadim Kaplonovsky, Serguei Krasnikov, Juan Maldecena, Aaron Bergman, and  Uday Varadarajan for helpful discussions.

\end{document}